\documentclass[12 pt, amsfonts, amssymb,color]{article}

\usepackage{amsmath,amssymb,amsfonts,latexsym,float,graphics,epsfig}
\usepackage{subfig}
\usepackage{verbatim}

\begin{document}

\renewcommand\theequation{\arabic{section}.\arabic{equation}}
\catcode`@=11 \@addtoreset{equation}{section}
\newtheorem{axiom}{Definition}[section]
\newtheorem{theorem}{Theorem}[section]
\newtheorem{axiom2}{Example}[section]
\newtheorem{lem}{Lemma}[section]
\newtheorem{prop}{Proposition}[section]
\newtheorem{cor}{Corollary}[section]
\newcommand{\be}{\begin{equation}}
\newcommand{\ee}{\end{equation}}

\newcommand{\equal}{\!\!\!&=&\!\!\!}
\newcommand{\rd}{\partial}
\newcommand{\g}{\hat {\cal G}}
\newcommand{\bo}{\bigodot}
\newcommand{\res}{\mathop{\mbox{\rm res}}}
\newcommand{\diag}{\mathop{\mbox{\rm diag}}}
\newcommand{\Tr}{\mathop{\mbox{\rm Tr}}}
\newcommand{\const}{\mbox{\rm const.}\;}
\newcommand{\leqnomode}{\tagsleft@true\let\veqno\@@leqno}
\newcommand{\reqnomode}{\tagsleft@false\let\veqno\@@eqno}

\newcommand{\cA}{{\cal A}}
\newcommand{\bA}{{\bf A}}
\newcommand{\Abar}{{\bar{A}}}
\newcommand{\cAbar}{{\bar{\cA}}}
\newcommand{\bAbar}{{\bar{\bA}}}
\newcommand{\cB}{{\cal B}}
\newcommand{\bB}{{\bf B}}
\newcommand{\Bbar}{{\bar{B}}}
\newcommand{\cBbar}{{\bar{\cB}}}
\newcommand{\bBbar}{{\bar{\bB}}}
\newcommand{\bC}{{\bf C}}
\newcommand{\cbar}{{\bar{c}}}
\newcommand{\Cbar}{{\bar{C}}}
\newcommand{\Hbar}{{\bar{H}}}
\newcommand{\bL}{{\bf L}}
\newcommand{\Lbar}{{\bar{L}}}
\newcommand{\cLbar}{{\bar{\cL}}}
\newcommand{\bLbar}{{\bar{\bL}}}
\newcommand{\cM}{{\cal M}}
\newcommand{\bM}{{\bf M}}
\newcommand{\Mbar}{{\bar{M}}}
\newcommand{\cMbar}{{\bar{\cM}}}
\newcommand{\bMbar}{{\bar{\bM}}}
\newcommand{\cP}{{\cal P}}
\newcommand{\cQ}{{\cal Q}}
\newcommand{\bU}{{\bf U}}
\newcommand{\bR}{{\bf R}}
\newcommand{\cW}{{\cal W}}
\newcommand{\bW}{{\bf W}}
\newcommand{\bZ}{{\bf Z}}
\newcommand{\Wbar}{{\bar{W}}}
\newcommand{\Xbar}{{\bar{X}}}
\newcommand{\cWbar}{{\bar{\cW}}}
\newcommand{\bWbar}{{\bar{\bW}}}
\newcommand{\abar}{{\bar{a}}}
\newcommand{\nbar}{{\bar{n}}}
\newcommand{\pbar}{{\bar{p}}}
\newcommand{\tbar}{{\bar{t}}}
\newcommand{\ubar}{{\bar{u}}}
\newcommand{\utilde}{\tilde{u}}
\newcommand{\vbar}{{\bar{v}}}
\newcommand{\wbar}{{\bar{w}}}
\newcommand{\phibar}{{\bar{\phi}}}
\newcommand{\Psibar}{{\bar{\Psi}}}
\newcommand{\bLambda}{{\bf \Lambda}}
\newcommand{\bDelta}{{\bf \Delta}}
\newcommand{\p}{\partial}
\newcommand{\om}{{\Omega \cal G}}
\newcommand{\ID}{{\mathbb{D}}}
\newcommand{\pr}{{\prime}}
\newcommand{\prr}{{\prime\prime}}
\newcommand{\prrr}{{\prime\prime\prime}}
\title{ Quadratically damped oscillators with non-linear restoring force}
\author{Ankan Pandey$^1$\footnote{E-mail: ankan.pandey@bose.res.in}, 
A Ghose Choudhury$^2$\footnote{E-mail aghosechoudhury@gmail.com}, 
Partha Guha$^{1}$\footnote{E-mail: partha@bose.res.in}\\
\\
$^1$ S N Bose National Centre for Basic Sciences\\ 
JD Block, Sector III, Salt Lake, Kolkata 700106,  India\\
$^2$ Department of Physics, Surendranath  College\\ 24/2 Mahatma
Gandhi Road, Calcutta 700009, India\\}

\date{ }

 \maketitle

\smallskip

\begin{abstract}
\textit{ In this paper we qualitatively analyse quadratically damped oscillators with non-linear restoring force. In particular, we obtain Hamiltonian structure and analytical form of the energy functions.}
\end{abstract}

\smallskip

\section{Introduction}

Modelling of physical systems is one of the key goals of mathematical
physics. Physical systems, in reality, are not ideal and systems like,
the harmonic oscillator and the viscously damped oscillators are actually idealisations
of real life systems. Real life systems are generally non-linear in character and are
usually represented by a variety of different mathematical descriptions.
To model the phenomenon of damping, specially at higher velocities it is necessary to introduce a quadratic
non-linearity and the corresponding term is in general of the form - $sgn(\dot{x})\dot{x}^{2}$ {\cite{NM,RD}}.
In recent times, such systems have gained much attention because of vast
range of applications in engineering, specially in the
problems involving hydrological drag and aerodynamics {\cite{YN,Mad}}. 

In {\cite{KR}}, Kovacic and Rakaric have considered a system where the quadratic damping coefficient
is a constant and with a non-linear forcing term of the form, $F(x)=sgn(x)|x|^{\alpha}$,
where $\alpha$ is positive real constant. For linear, i.e, viscous, damping the
systems can easily be solved using standard well established mathematical
techniques. However, with quadratic damping the complexity in solving
such problems using analytic methods increases. In the case of quadratic damping
with non-linear forcing term $F(x)$ the orbits involve
discontinuous jumps at every quadrant boundary. To exactly solve such
systems is not easy as four different systems have to be solved for
each quadrant and their solutions need to be matched at the respective boundaries. Numerical approaches are, therefore, necessary to deal
with such systems as they provide valuable insights of the qualitative
nature of the system.

Quadratic damping arise generally in flows with high flow velocity. In general, the drag force consists of linear as well as quadratic terms. For slow flows, the linear term dominates and at higher velocity the linear term becomes negligible and quadratic term dictates the motion. In fluid dynamics, Morison's equation  is one of the fundamental equations which gives the net in-line force on a fixed body in oscillatory flows. It is a semi-empirical equation and is given as 
\begin{equation}
F=a\dot{v}+bv|v|.
\end{equation}

 Here $a$ and $b$ are proportionality constants which depend upon the density and the geometry of the body and $v(t)$ is the flow velocity. The first term is the inertial force and the second term gives the drag force on the object. In \cite{YN}, Yamamoto and Nath studied the oscillatory flows around a cylinder and they determined the drag force as being directly proportional to the signed velocity squared where the coefficients were determined experimentally. In \cite{Mad}, Madison used a quadratic damping force caused by fluid friction to study isenthalpic oscillations in a saturated two-phase fluid. 

In this work, we consider a general system, 
\begin{equation}
x''+sgn(x')f(x)x'^{2}+sgn(x)|g(x)|=0,\label{eq:0.1}
\end{equation}
where the damping coefficient is a function of the displacement and study it
for different cases using different instances of $f(x)$ and $g(x)$.
We formulate a generalised scheme for the calculation of the mechanical energy given
by a undamped oscillator. We also derive Hamiltonian description for
the system which is basically piecewise in different quadrants.

The organisation of the paper is as follows. In section 2, we transform
the second-order differential equation (\ref{eq:0.1}) into a first-order differential equation
for the kinetic energy of the corresponding undamped oscillator. Further we give the Hamiltonian
structure for the system. To analyse the system, we consider
the specific cases of odd and even forcing terms and provide some illustrative examples. Finally
analytic expressions are calculated for the energy and the Hamiltonians.

\section{Oscillators with non-negative real-power restoring force and quadratic
damping.}

Oscillators with quadratic damping together with non-linear non-negative
real-power restoring force are interesting problems and were well studied
by I. Kovacic and Z. Rakaric in {\cite{KR}}, they consider a second-order differential equation of the form
\begin{equation}
x''+\mu\,sgn(x')x'^{2}+sgn(x)|x|^{\alpha}=0,\label{eq:1.1}
\end{equation}
where the prime denotes differentiation with respect to $t$, $\mu$
denotes the damping coefficient and $\alpha\in\mathbb{R}^{+}$. We study the above system in a more general setting with displacement
dependent damping so that $\mu=f(x)$, and with a forcing term $sgn(x)|g(x)|$.
The corresponding equation is then given by
\begin{equation}
x''+sgn(x')f(x)x'^{2}+sgn(x)|g(x)|=0.\label{eq:1.2}
\end{equation}

From (\ref{eq:1.2}) we have
\begin{equation}
\frac{d}{dt}\left[\frac{1}{2}(x')^{2}+\int_{0}^{x}|g(s)|\,ds\right]=-sgn(x')f(x)x'^{3}.\label{eq:1.3}
\end{equation}
Let $T=\frac{1}{2}(x')^{2}$, and,$V=\int_{0}^{x}|g(s)|\,ds$, then $E=T+V$.
Now, from (\ref{eq:1.3}) we can have
\begin{equation}
\frac{dT}{dx}+2\,sgn(x')f(x)\,T=-sgn(x)|g(x)|.\label{eq:1.4}
\end{equation}
In order to understand the qualitative features of the system we consider
the following two cases depending on the whether the forcing term
is even or odd.

\begin{figure}
\hfill{}\subfloat[$f(x)=x,\,\,g(x)=x^{3}$.]{\includegraphics[width=5cm,height=5cm,keepaspectratio]{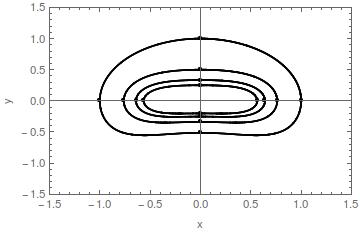}\label{fig:1a}}\hfill{}\subfloat[$f(x)=1,\,\,g(x)=x^{3}$.]{\includegraphics[width=5cm,height=5cm,keepaspectratio]{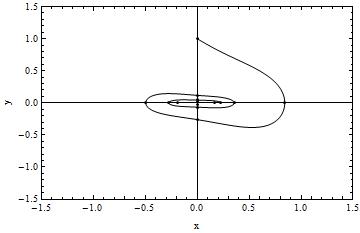}\label{fig:1b}}\hfill{}

\caption{For odd-power restoring force with odd (a) and even (b) damping coefficients. }
\label{fig:1}

\end{figure}

\begin{description}
\item [{Case~1:~odd-power~restoring~forces}]~
\end{description}
In this situation
(\ref{eq:1.2}) can be written as
\begin{equation}
x''+sgn(x')f(x)x'^{2}+g(x)=0,\label{eq:1.7}
\end{equation}
which is the same system considered in {\cite{PGP}}. As shown in {\cite{PGP}} for
even damping coefficient, we have a damped orbit (fig(\ref{fig:1b}))
and for odd damping we have a closed orbit (fig(\ref{fig:1a})).

\begin{figure}
\hfill{}\subfloat[$f(x)=1,\,g(x)=x^{3}$.]{\includegraphics[width=5cm,height=5cm,keepaspectratio]{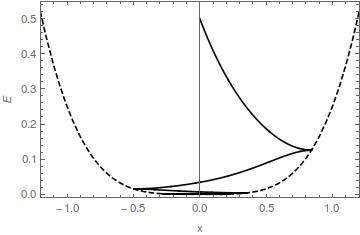}\label{fig:2a}

}\hfill{}\subfloat[$f(x)=x,\,g(x)=x^{3}$.]{\includegraphics[width=5cm,height=5cm,keepaspectratio]{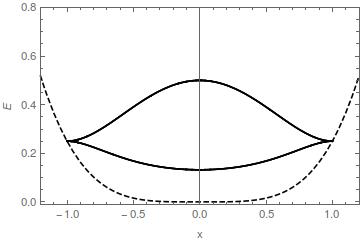}\label{fig:2b}

}\hfill{}

\caption{Energy variation with displacement for odd-power restoring forces.}
\label{fig:2}

\end{figure}

As an example when $f(x)=1,\,g(x)=x^{3}$, the system is given as
\leqnomode
\[
 x''+sgn(x')x'^{2}+x^{3}=0\,\tag{\ensuremath{\textbf{example\,1}}}
\]
\reqnomode
Then from (\ref{eq:1.4}) we have
\[
\frac{dT}{dx}+2T=-x^{3}\,\,\,\,\,\,x'>0\tag{2.16a},
\]
\[
\frac{dT}{dx}-2T=-x^{3}\,\,\,\,\,\,x'<0\tag{2.16b}.
\]
Solution to the above equations are given as
\[
T=\frac{1}{8}\left(3-6x+6x^{2}-4x^{3}\right)+c_{+}\,e^{-2x}\,\,\,\,\,x'>0,
\]
\[
T=\frac{1}{8}\left(3+6x+6x^{2}+4x^{3}\right)+c_{-}\,e^{2x}\,\,\,\,\,x'<0,
\]
where $c_{\pm}$ are constants of integration and are fixed using
the initial conditions. The corresponding energies are
\[
E=T+V=\frac{1}{8}\left(3-6x+6x^{2}-4x^{3}\right)+c_{+}\,e^{-2x}+\frac{x^{4}}{4},\,\,\,\,x'>0,
\]
\[
E=T+V=\frac{1}{8}\left(3+6x+6x^{2}+4x^{3}\right)+c_{-}\,e^{2x}+\frac{x^{4}}{4},\,\,\,\,x'<0.
\]
Now consider another example with odd damping coefficient, $f(x)=x$.
In this setting, (\ref{eq:1.7}) becomes 
\leqnomode
\[
x''+sgn(x')xx'^{2}+x^{3}=0\,\tag{\ensuremath{\textbf{example\,2}}}
\]
\reqnomode
and (\ref{eq:1.4}) takes the form
\[
\frac{dT}{dx}+2xT=-x^{3}\,\,\,\,\,\,x'>0\tag{2.17a},
\]
\[
\frac{dT}{dx}-2xT=-x^{3}\,\,\,\,\,\,x'<0\tag{2.17b}.
\]
The final energy is given as
\[
E=T+V=\frac{1}{2}\left(1-x^{2}\right)+c_{+}\,e^{-x^{2}}+\frac{x^{4}}{4},\,\,\,\,x'>0,
\]
\[
E=T+V=\frac{1}{2}\left(1+x^{2}\right)+c_{-}\,e^{x^{2}}+\frac{x^{4}}{4},\,\,\,\,x'<0,
\]
where $c_{\pm}$ are constants of integration. Figure (\ref{fig:2a},
\ref{fig:2b}) shows the energy-displacement curve for the above considered
examples, and the corresponding phase space orbits are shown in figure (\ref{fig:1a},
\ref{fig:1b}).

\begin{figure}
\hfill{}\subfloat[$f(x)=x,\,\,g(x)=x^{2}$.]{\includegraphics[width=5cm,height=5cm,keepaspectratio]{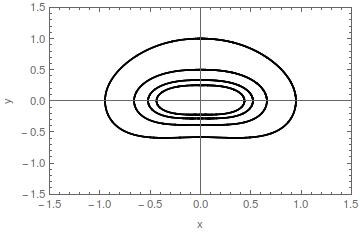}\label{fig:3a}

}\hfill{}\subfloat[$f(x)=1,\,\,g(x)=x^{2}$.]{\includegraphics[width=5cm,height=5cm,keepaspectratio]{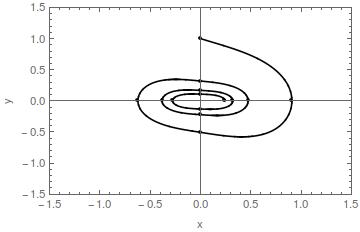}\label{fig:3b}

}\hfill{}

\caption{Even-power restoring with odd (a) and even (b) damping.}
\label{fig:3}

\end{figure}

\begin{description}
\item [{Case~2:~even-power~restoring~forces}]~
\end{description}
In the case of even-power restoring force, the forcing term is again
an odd function and behaves qualitatively in the same manner as in the odd-power
case. However, the damping and the period will vary depending upon
$\alpha$. Figure(\ref{fig:3}) shows the orbits for even and
odd damping coefficients.

\begin{figure}

\hfill{}\subfloat[$f(x)=1,\,g(x)=x^{2}$.]{

\includegraphics[width=5cm,height=5cm,keepaspectratio]{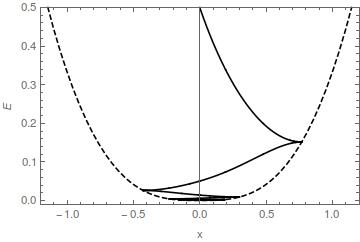}\label{fig:4a}

}\hfill{}\subfloat[$f(x)=x,\,g(x)=x^{2}$.]{

\includegraphics[width=5cm,height=5cm,keepaspectratio]{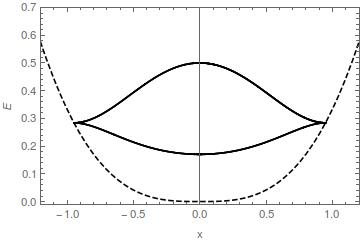}\label{fig:4b}

}\hfill{}

\caption{Energy-displacement curve for even and odd damping functions. }
\label{fig:4}

\end{figure}

The equations for this case are given as
\[
x''+sgn(x')f(x)x'^{2}+g(x)=0,\,\,\,\,\,\,x<0,\tag{2.18a}
\]
\[
x''+sgn(x')f(x)x'^{2}-g(x)=0,\,\,\,\,\,\,x<0.\tag{2.18b}
\]
The dynamics in this case varies in each quadrants of the phase plane.
To illustrate this case, consider an example with $f(x)=1,\,g(x)=x^{2}$.
The system is given as
\leqnomode
\[
x''+sgn(x')x'^{2}+sign(x)x^{2}=0.\,\tag{\ensuremath{\textbf{example\,3}}}
\]
\reqnomode
The dependence of energy on displacement can be computed using (\ref{eq:1.4})
and $E=T+V$ and leads to the following expressions

\[
E=\frac{1}{4}\left(-1+2x-2x^{2}\right)+c_{++}e^{-2x}+\frac{x^{3}}{3},\,\,\,\,\,x>0,\,x'>0
\]
\[
E=\frac{1}{4}\left(1+2x+2x^{2}\right)+c_{-+}e^{2x}+\frac{x^{3}}{3},\,\,\,\,\,x<0,\,x'>0
\]
\[
E=\frac{1}{4}\left(1-2x+2x^{2}\right)+c_{+-}e^{-2x}+\left|\frac{x^{3}}{3}\right|,\,\,\,\,\,x>0,\,x'<0
\]
\[
E=\frac{1}{4}\left(-1-2x-2x^{2}\right)+c_{--}e^{2x}+\left|\frac{x^{3}}{3}\right|,\,\,\,\,\,x<0,\,x'<0
\]
where $c_{\pm\pm}$ are constants of integration. Figure (\ref{fig:4a})
shows the energy variation with displacement with the  dotted lines denoting the
potential $V$.

For an example with closed orbits, assume $f(x)=x$. The
system is given as
\leqnomode
\[
x''+sgn(x')xx'^{2}+sign(x)x^{2}=0.\,\tag{\ensuremath{\textbf{example\,4}}}
\]
\reqnomode
By following a similar procedure as outlined before we find the energy dependence on displacement to be given by
\[
E=e^{-x^{2}}\left(-\frac{1}{2}e^{x^{2}}x+\frac{1}{4}\sqrt{\pi}\mathcal{E}rfi(x)\right)+c_{++}e^{-x^{2}}+\frac{x^{3}}{3},\,\,\,\,\,x>0,\,x'>0
\]
\[
E=e^{x^{2}}\left(\frac{1}{2}e^{-x^{2}}x-\frac{1}{4}\sqrt{\pi}\mathcal{E}rf(x)\right)+c_{-+}e^{x^{2}}+\frac{x^{3}}{3},\,\,\,\,\,x<0,\,x'>0
\]
\[
E=e^{-x^{2}}\left(\frac{1}{2}e^{x^{2}}x-\frac{1}{4}\sqrt{\pi}\mathcal{E}rfi(x)\right)+c_{+-}e^{-x^{2}}+\left|\frac{x^{3}}{3}\right|,\,\,\,\,\,x>0,\,x'<0
\]
\[
E=e^{x^{2}}\left(-\frac{1}{2}e^{-x^{2}}x+\frac{1}{4}\sqrt{\pi}\mathcal{E}rf(x)\right)+c_{--}e^{x^{2}}+\left|\frac{x^{3}}{3}\right|,\,\,\,\,\,x<0,\,x'<0
\]
where $c_{\pm\pm}$ are constants of integration and $\mathcal{E}rf\,and\,\mathcal{E}rfi$
are error functions and imaginary error functions. Figure (\ref{fig:4b})
shows the energy variation with displacement with dotted lines denoting
potential $V$.

\subsection{The Hamiltonian structure of the system}

In the above analysis the mechanical energy $E=T+V$ corresponds to that of a particle of unit mass moving in the potential $V$. Owing to the presence of the damping term it is quite natural that the mechanical energy is not conserved. However, it is interesting to note that in the case of ODEs of the form of (\ref{eq:0.1}) one can formulate an alternative description in terms of a variable mass. In the following we consider first a second-order ODE with a quadratic dependence on the velocity given by
\be\label{E1} \ddot{x}+f(x)\dot{x}^2 +g(x)=0.\ee
we assume $f(x) $ and $g(x)$ are such that $f(0)=g(0)=0$ and $f(x)$ is integrable while $g^\prime(0)>0$. The functional form of $g(x)=g^\prime(0) x +g_n(x)$ where $g_n(x)$ is analytic. As demonstrated in \cite{NL3,NUT}, the Jacobi Last Multiplier (JLM) provides a convenient tool for obtaining a Lagrangian for second-order equations of the form $\ddot{x}=\mathcal{F}(x, \dot{x})$. It is defined as a solution of
\be\label{E2} \frac{d}{dt} \log M+\frac{\partial \mathcal{F}(x, \dot{x})}{\partial \dot{x}}=0.\ee
In the present case it follows that
\be\label{E3} M=\exp\left(2F(x)\right), \;\;\;\; \hbox{ where }  \;\;\;\;  F(x)=\int_0^x f(s) ds.\ee
The relationship between the JLM, $M$ and the Lagrangian is provided by $M=\partial^2 L/\partial \dot{x}^2$ as a consequence of which for (\ref{E1})  the Lagrangian may be expressed in the form
\be\label{E4} L=\frac{1}{2}e^{2F(x)} \dot{x}^2 -V(x),\ee
where $V(x)$ is determined by substituting (\ref{E4}) into the Euler-Lagrange equation $$\frac{d}{dt}\left(\frac{\partial L}{\partial \dot{x}}\right)=\left(\frac{\partial L}{\partial x}\right),$$ which immediately gives
\be\label{E5} V(x)=\int_0^x e^{2F(s)} g(s) ds.\ee
By means of the standard Legendre transformation we can obtain the Hamiltonian as
\be\label{E6}H=\frac{1}{2}e^{2F(x)} \dot{x}^2+\int_0^x e^{2F(s)} g(s) ds.\ee It is easily verified that the Hamiltonian is a constant of motion and the expression for the conjugate momentum $ p=e^{2F(x)} \dot{x}$  suggests that $M=e^{2F(x)}$ serves as a position dependent mass term. In fact equations with quadratic velocity dependance of the type considered here naturally arise in the Newtonian formulation of the equation of motion of a particle with a variable mass.
Clearly then, the trajectories for arbitrary intial condition $(x_0, y_0)$ where $ y=\dot{x}$ are given by
\be\label{eq:E7} \frac{1}{2}e^{2F(x)} y^2+V(x)=\frac{1}{2}e^{2F(x)} y_0^2+V(x_0).\ee

\bigskip

In terms of the canonical momentum, $p=e^{2F(x)} \dot{x}$, the Hamiltonian $H$ becomes 
\be H = \frac{p^2}{2e^{2F(x)}} + V(x).
\ee
From the above it is obvious that in the case of oscillators with quadratic damping and normal non-linear
forcing terms, the Hamiltonian for system (\ref{eq:1.1}) is given by 
\begin{equation}
H^{\pm\pm}=\frac{1}{2}e^{\pm2F(x)}x'^{2}\pm\int_{0}^{x}e^{\pm2F(s)}g(s)\,ds,\label{eq:1.5}
\end{equation}
where $F(x)=\int_{0}^{x}f(s)\,ds$ and the first $(\pm)$ in the superscript
of $H$ denotes $(\pm)$ in the power of exponential which basically
comes from sign function in the coefficient of damping term while
the second one denotes $(\pm)$ in the forcing term. It is also natural to define the potential function by
\begin{equation}
V_{H}^{\pm}=\int_{0}^{x}e^{\pm2F(s)}g(s)\,ds.\label{eq:1.6}
\end{equation}
 In the following we provide explicit expressions for the Hamiltonian for our previous examples and also list the comparative changes in the values of $H$ and $E$ for each case.
The Hamiltonian for the above examples are given as

$example\,:\,f(x)=1,\,g(x)=x^{3}$

\[
H^{\pm}=\frac{1}{2}e^{\pm2x}x'^{2}+\frac{(3+e^{\pm2x}\left(-3\pm2x\,(3\pm x(-3\pm2x)))\right)}{8},
\]

$example\,:\,f(x)=x,\,g(x)=x^{3}$
\[
H^{\pm}=\frac{1}{2}e^{\pm x^{2}}x'^{2}+\frac{\left(1+e^{\pm x^{2}}(-1+\pm x^{2})\right)}{2}.
\]

The above Hamiltonians  do not depend on the sign of displacement. This
is evident from (\ref{eq:1.7}). Table (\ref{tab:1}) shows the changes
in the values of the Hamiltonian and energy $E$ in each complete cycle.

\begin{table}
\hfill{}%
\begin{tabular}{|c|c|c|}
\hline 
$n^{th}$ cycle & $\Delta H$ & $\Delta E$\tabularnewline
\hline 
\hline 
1 & -0.493039 & -0.493039\tabularnewline
\hline 
2 & -0.00603154 & -0.00603154\tabularnewline
\hline 
3 & -0.000307447 & -0.000684351\tabularnewline
\hline 
4 & -0.0000497198 & -0.000154833\tabularnewline
\hline 
\end{tabular}\hfill{}

\caption{Change in Hamiltonian and Energy in each full cycle.}
\label{tab:1}

\end{table}

The Hamiltonian for the examples in even-power case are given as
\[
H^{s\,w}=\frac{1}{2}e^{s\,x^{2}}x'^{2}+\frac{w\left(2e^{s\,x^{2}}\sqrt{s}x-\sqrt{\pi}\mathcal{E}rfi(\sqrt{s}x)\right)}{4s^{3/2}},
\]
where $s,\,w=\{\pm1\}$ and denotes the sign of displacement $x$
and velocity $x'$, respectively. Table (\ref{tab:2}) shows the change
in the values of Hamiltonian and energy in each cycle.

\begin{table}
\hfill{}%
\begin{tabular}{|c|c|c|}
\hline 
$n^{th}$ cycle & $\Delta H$ & $\Delta E$\tabularnewline
\hline 
\hline 
1 & -0.486074 & -0.486074\tabularnewline
\hline 
2 & -0.011064 & -0.0278193\tabularnewline
\hline 
3 & -0.000956281 & -0.000956281\tabularnewline
\hline 
4 & -0.000216832 & -0.000714582\tabularnewline
\hline 
\end{tabular}\hfill{}

\caption{Change in Hamiltonian and energy in each complete cycle.}
\label{tab:2}

\end{table}

\section{Conclusion}

In this paper, a general quadratically damped oscillator with non-linear force is studied. The associated energy, taken as energy of an undamped oscillator, has been calculated and an analytic expression for the same is obtained. Hamiltonian description of the system is shown and analytic expression has been obtained for different instances.

\end{document}